\documentclass[aip,jmp,reprint]{revtex4-1}

\usepackage{import}
\usepackage{dcolumn}
\usepackage{amsmath}  
\usepackage{amsfonts} 
\usepackage{textcomp}
\usepackage[utf8]{inputenc} 
\usepackage[T1]{fontenc}    
\usepackage{hyperref}       
\usepackage{url}            
\usepackage{booktabs}       
\usepackage{nicefrac}       
\usepackage{microtype}      
\usepackage{lipsum}
\usepackage[pdftex]{graphicx}
\usepackage{caption}
\usepackage{amssymb, mathtools, amsthm}
\usepackage{xcolor}
\usepackage{mathrsfs}
\usepackage{tensor}
\usepackage{todonotes}
\usepackage{float} 
\usepackage{ragged2e}
\usepackage{scalerel}
\usepackage[normalem]{ulem}
\usepackage[english]{babel}
\usepackage{graphicx}
\usepackage{dcolumn}
\usepackage{bm}
\usepackage{blindtext}
\usepackage{verbatim}
\usepackage{relsize}
\usepackage{musicography}
\usepackage{blindtext}
\usepackage{cancel}
\usepackage{physics}
\usepackage{epstopdf}
\usepackage{mathtools}
\usepackage{blindtext}
\usepackage{color}
\usepackage[usenames,dvipsnames]{pstricks}
\usepackage{epsfig}
\usepackage{pst-grad} 
\usepackage{pst-plot} 
\usepackage{hyperref}
\usepackage{verbatim}
\usepackage{slashed}
\usepackage{subcaption}
\usepackage{dsfont}
\usepackage[english]{babel}
\usepackage{cleveref}

\newcommand{\p}{\partial}

\newcommand{\CC}{\mathbb{C}}
\newcommand{\RR}{\mathbb{R}}

\newcommand{\sop}{\mathrm{SO}^+}

\newcommand{\isop}{\mathrm{ISO}^+}
\newcommand{\id}{\mathds{1}}

\newcommand{\ii}{\mathrm{i}}

\newcommand{\xx}{\mathsf{x}}

\usepackage{float}
\usepackage{tikz-cd}
\usepackage{amsthm}
\usepackage{geometry}
 \usepackage{braket}

\theoremstyle{plain}

\theoremstyle{definition}

\usepackage[final]{showlabels} 

\newcommand{\mrm}[1]{\mathrm{#1}}

\newcommand{\mfk}[1]{\mathfrak{#1}}

\newcommand{\beq}{\begin{equation}}
\newcommand{\eeq}{\end{equation}}

\newcommand{\ba}{\begin{align}}
\newcommand{\ea}{\end{align}}

\allowdisplaybreaks[1] 

\draft 

\begin{document}

\title{Stationary trajectories in Minkowski spacetimes} 

\author{Cameron R D Bunney}
\email[]{cameron.bunney@nottingham.ac.uk}
\affiliation{School of Mathematical Sciences, University of Nottingham, Nottingham NG$7$ $2$RD, UK}

\date{October 2023; revised April 2024 \\ aaPublished in \textit{J}.\ \textit{Math}.\ \textit{Phys}.\ \textbf{65}, 052501 (2024), doi.org/10.1063/5.0205471\\ aaFor Open Access purposes, this Author Accepted Manuscript is made available under CC BY public copyright.}

\begin{abstract}
    We determine the conjugacy classes of the Poincaré group $\isop(n,1)$ and apply this to classify the stationary trajectories of Minkowski spacetimes in terms of timelike Killing vectors. Stationary trajectories are the orbits of timelike Killing vectors and, equivalently, the solutions to Frenet-Serret equations with constant curvature coefficients. We extend the $3+1$ Minkowski spacetime Frenet-Serret equations due to Letaw to Minkowski spacetimes of arbitrary dimension. We present the explicit families of stationary trajectories in $4+1$ Minkowski spacetime.
\end{abstract}


\maketitle 

\section{Introduction}
Stationary trajectories in Minkowski spacetime can be defined as the timelike solutions to the Frenet-Serret equations, whose curvature invariants are constant in proper time. Letaw~\cite{Letaw} showed in the case of $3+1$ Minkowski spacetime that these trajectories equivalently correspond to the orbits of timelike Killing vectors.

The study of curves defined in terms of their curvature invariants began with the work of Frenet~\cite{Frenet1852} and Serret~\cite{Serret1851} in three-dimensional, flat, Euclidean-signature space with the standard metric. Frenet and Serret found a coupled set of differential equations for the tangent vector, normal vector, and binormal vector, which together form an orthonormal basis in $\RR^3$. These differential equations are now known as the Frenet-Serret equations and were later generalised by Jordan~\cite{Jordan} to flat, Euclidean spaces of arbitrary dimension.

We are interested in the stationary trajectories of Minkowski spacetimes. One motivation in physics where stationary trajectories are extensively used is in the study of the Unruh effect~\cite{Unruh} and Unruh-like effects~\cite{Biermann, BunneyThermal, Good:2020hav, Fewster, Takagi}. They have the exploitable property that, in quantum field theory, the two-point function with respect to the Minkowski vacuum when pulled back to a stationary worldline is only a function of the total proper time~\cite{Biermann}. As such, the response of a particle detector~\cite{Unruh,DeWitt:1980hx} is time independent~\cite{Fewster} and hence there is no time dependence in the detector's associated spectrum~\cite{Letaw}.

The purpose of this paper is to classify the stationary trajectories of Minkowski spacetimes of any dimension. Because stationary trajectories can be defined in terms of curves with constant curvature invariants and alternatively in terms of timelike Killing vectors, we develop the classification in both of these formalisms. First, we determine the conjugacy classes of the connected component of the Poincaré group, referred to as the restricted Poincaré group, and second, we extend the Frenet-Serret equations in $3+1$ Minkowski spacetime to $n+1$ Minkowki spacetime. Related previous work includes the classification of the adjoint and co-adjoint orbits of the Poincare\'e group~\cite{BURGOYNE1977339,cushman2006adjoint}. A classification of the stationary trajectories in terms of the eigenvalues of a matrix constructed from the Killing vector was given in~\cite{townsend_stationary}, with explicit results up to six spacetime dimensions.

We begin in Section~\ref{sec:: KV minkowski} with the first method by reviewing the isometries of Minkowski spacetime in terms of the restricted Poincaré group $\isop(n,1)$ where $n+1$ is the dimension of Minkowki spacetime. We then classify the conjugacy classes of the restricted Lorentz group $\sop(n,1)$ in Section~\ref{subsec:: lorentz conj} by generalising the known results for $\sop(3,1)$. In Section~\ref{subsec:: poincare conj}, we extend the classification to fully determine the conjugacy classes of the Poincaré group. 

Specialising to the conjugacy classes whose associated Killing vector is timelike somewhere, we classify the stationary trajectories in $n+1$ Minkowski spacetime in Section~\ref{subsec:: stationary traj}, presenting the set of all timelike Killing vectors~\eqref{eqn:: TKV set} and giving a formula for the number of classes of timelike trajectories~\eqref{eqn:: TKV size}.

In Section~\ref{sec:: tetrad formulation}, we consider the second method and extend the Frenet-Serret equations of $3+1$ Minkowski spacetime to $n+1$ Minkowski spacetime. We present the ordinary differential equation satisfied by the four-velocities of the stationary trajectories. Finally, in Section~\ref{sec:: 4+1 examples}, we use this formalism to explicitly present the stationary trajectories of $4+1$ Minkowski spacetime, showing that these trajectories fall into nine distinct families.

We use units in which the speed of light is set to unity. Sans serif letters ($\xx$) denote spacetime points and boldface Italic letters ($\bm{x}$) denote spatial vectors.  We adopt the mostly plus convention for the metric of Minkowski spacetime $\dd s^2=-\dd t^2+\dd\bm{x}^2$ and use the standard set of Minkowski coordinates $(t,x,y,z,\dots)$.

\section{Killing vectors in Minkowski spacetimes}\label{sec:: KV minkowski}

Stationary trajectories are the timelike solutions to the Frenet-Serret equations with proper-time-independent curvature invariants. Letaw~\cite{Letaw} demonstrated that the stationary trajectories of $3+1$ Minkowski spacetime can be alternatively defined as the orbits of timelike Killing vectors. Each solution to the Frenet-Serret equations is determined only up to a Poincaré transformation of the worldline, leading to equivalence classes of trajectories. In terms of Killing vectors, each is determined up to conjugation of the generator associated to the Killing vector. In this Section, we determine the conjugacy classes of the Poincaré group, then restrict to the classes whose associated Killing vector is timelike somewhere, thereby classifying the stationary trajectories of Minkowski spacetime of dimension $n+1$, where $n\geq 1$.

\subsection{Isometries of Minkowski spacetime}\label{subsec:: minkowski isometries}
The isometry group of Minkowski spacetime $\RR^{n,1}$ is the Poincaré group. We consider only the connected component of the Poincaré group, the restricted Poincaré group $\mrm{ISO}^+(n,1)$, consisting of the connected component of the Lorentz group and translations. A general element of $\mrm{ISO}^+(n,1)$ is a pair $g=(\Lambda,a)$, where $\Lambda$ is an element of the restricted Lorentz group $\mrm{SO}^+(n,1)$ and $a\in\RR^{n,1}$. The restricted Lorentz group is the subgroup of Lorentz transformations preserving orientation and time orientation. The Poincaré group acts on $\RR^{n,1}$ by $g\,\xx^\mu={\Lambda^\mu}_\nu\xx^\nu+a^\nu$. The Poincaré group is equipped with the group multiplication law $\widetilde{g}\cdot g=(\widetilde{\Lambda},\widetilde{a})\cdot(\Lambda,a)=(\widetilde{\Lambda}\Lambda,\widetilde{\Lambda}a+\widetilde{a})$ and inverse elements are given by $g^{-1}=(\Lambda^{-1},-\Lambda^{-1}a)$. The restricted Lorentz group is a subgroup of the restricted Poincaré group with elements $(\Lambda,0)$. Pure spacetime translations $h=(\id,a)$ form a normal subgroup of the Poincaré group, which may be verified by explicitly computing $g\cdot h\cdot g^{-1}$. As such, this decomposes $\isop(n,1)$ as a semidirect product, $\mrm{ISO}^+(n,1)=\RR^{n,1}\rtimes\mrm{SO}^+(n,1)$. This structure as a semidirect product of Lie groups is inherited at the level of Lie algebras, $\mfk{iso}(n,1)=\RR^{n,1}\rtimes\mfk{so}(n,1)$.

Given a Lie group $G$ and associated Lie algebra $\mfk{g}$, $G$ acts naturally on $\mfk{g}$ by conjugation, $G\times\mfk{g}\rightarrow\mfk{g}$, $(g,X)\mapsto gXg^{-1}$. We use a matrix group notation, anticipating its use in Section~\ref{subsec:: poincare conj}. We define a conjugacy class in the Lie algebra in the following sense $Y\sim X \iff \exists g\in G$ such that $Y=gXg^{-1}$.

When considering the Poincaré group acting on Minkowski spacetime, the generators of the Lie algebra $\mfk{iso}(n,1)$ are the Killing vector fields. A Killing vector field is the velocity vector field of a one-parameter isometry group at the identity. It is natural then to consider representations of these generators. The infinitesimal Poincaré transformation of a scalar field $\phi(\xx)$ leads to $\phi(\xx)\mapsto \phi(\xx)-(a^\mu\p_\mu+\tfrac{1}{2}\omega^{\mu\nu}(\xx_\nu\p_\mu-\xx_\mu\p_\nu))\phi(\xx)$, where $\omega^{\mu\nu}$ is antisymmetric. This is the standard vector field representation and may be written as $\phi(\xx)\mapsto(1-a^\mu P_\mu-\tfrac{1}{2}\omega^{\mu\nu}M_{\mu\nu})\phi(\xx)$, where $P_\mu=\p_\mu$ is the generator of spacetime translations and $M_{\mu\nu}=(\xx_\nu\p_\mu-\xx_\mu\p_\nu)$ is the generator of spacetime rotations.

Alternatively, Killing vectors $\xi=\xi^\mu\p_\mu$ of a spacetime $\mathcal{M}$ are defined by,
\begin{equation}\label{eqn:: Killing vectors1}
   \nabla_\mu\xi_\nu+\nabla_\nu\xi_\mu~=~0\,.
\end{equation} Combining~\eqref{eqn:: Killing vectors1} with the Ricci identity, one may show the following identity holds~\cite{Carroll:2004st}
\begin{equation}\label{eqn:: Killing vectors2}
    \nabla_\mu\nabla_\nu\xi_\sigma~=~{R^\rho}_{\mu\nu\sigma}\xi_\rho\,.
\end{equation}
In the case of Minkowski spacetime $\RR^{n,1}$ in Minkowski coordinates, equations~\eqref{eqn:: Killing vectors1} and~\eqref{eqn:: Killing vectors2} reduce to the following,
\begin{subequations}\label{subeqn:: Minkowski Killing vectors}
    \begin{align}
        \p_\mu\xi_\nu+\p_\nu\xi_\mu&~=~0\,,\label{eqn:: Minkowski Killing equation}\\
        \p_\mu\p_\nu\xi_\sigma&~=~0\,.\label{eqn:: Minkowski flat eqn}
    \end{align}
\end{subequations}
One may integrate~\eqref{eqn:: Minkowski flat eqn} and combine it with~\eqref{eqn:: Minkowski Killing equation} to write 
\begin{equation}\label{eqn:: general Killing vector}
    \xi^\mu~=~c^\mu+\omega^{\mu\nu}\xx_\nu\,,
\end{equation}
where $c^\mu$ is a constant and $\omega^{\mu\nu}$ is antisymmetric, $\omega^{\mu\nu}=-\omega^{\nu\mu}$, leading to $\xi=c^\mu\p_\mu+\tfrac{1}{2}\omega^{\mu\nu}(\xx_\nu\p_\mu-\xx_\mu\p_\nu)$.

With foresight, we define the following notation with $i<j$:
\begin{subequations}\label{eqn:: all Killing vectors}
    \begin{align}
        \mrm{T}_0&~=~\p_t\,,\\
        \mrm{S}_i&~=~\p_i\,,\\
        \mrm{NT}_{0i}&~=~\p_t-\p_i\,,\\
        \mrm{B}_{0i}&~=~t\p_i+x_i\p_t\,,\\
        \mrm{R}_{ij}&~=~x_j\p_i-x_i\p_j\,,\\
        \mrm{NR}_{0ij}&~=~\mrm{B}_{0i}-\mrm{R}_{ij}\,,\nonumber\\
        &~=~(t-x_j)\p_i+x_i(\p_t+\p_j)\,,
    \end{align}
\end{subequations}
where $\mrm{R}_{ij}$ is the Killing vector associated with a rotation in the $x^i$\nobreakdash--$x^j$ plane, $\mrm{NR}_{0ij}$ is the Killing vector associated with a null rotation consisting of a boost along the $x^i$\nobreakdash--axis and a rotation in the $x^i$\nobreakdash--$x^j$ plane, $\mrm{B}_{0i}$ is the Killing vector associated with a boost along the $x^i$\nobreakdash--axis, $\mrm{T}_0$ is the Killing vector associated with a timelike translation, $\mrm{S}_i$ is the Killing vector associated with a spacelike translation parallel to the $x^i$\nobreakdash--axis, and $\mrm{NT}_{0i}$ is the Killing vector associated with a null translation with spatial translation parallel to the $x^i$\nobreakdash--axis.
We use $\oplus$ as a shorthand for Killing vectors with scalars suppressed. For example, $\xi=\mrm{T}_0\oplus \mrm{R}_{12}$ is shorthand for $\xi=a\p_t+b(x_2\p_1-x_1\p_2)$ with $a$, $b$ both nonzero.

\subsection{Conjugacy classes of the restricted Lorentz group}\label{subsec:: lorentz conj}

$\isop(n,1)$ is a semidirect product of the restricted Lorentz group and the group of spacetime translations. Owing to this decomposition, one may methodically approach determining the conjugacy classes of the Poincaré group by first beginning with the restricted Lorentz group and then considering the effect of spacetime translations.

\subsubsection{Conjugacy classes of the Möbius group}

\noindent To classify the conjugacy classes of $\mrm{SO}^+(n,1)$, we first consider $\mrm{SO}^+(3,1)$, which is isomorphic to the Möbius group $\mrm{PSL}(2,\CC)=\mrm{SL}(2,\CC)/\{\id,-\id\}$ where $\id$ is the identity matrix. To see this, one notes that there is a homomorphism between $\RR^{3,1}$ and anti-hermitian matrices by sending $\xx^\mu$ to $\ii(\xx^0\id+\bm{x}\cdot\bm{\sigma})$, where $\bm{\sigma}$ are the Pauli matrices. The determinant of the resulting matrix is the Minkowski squared distance from the origin, $\xx^\mu\xx_\mu$. The special linear group $\mrm{SL}(2,\CC)$ acts naturally on the set of anti-hermitian matrices by conjugation, which preserves the determinant and hence preserves the Minkowski squared distance. This implies a (surjective) homomorphism $\mrm{SL}(2,\CC)\rightarrow \mrm{SO}^+(3,1)$. The kernel of this map is $\{\id,-\id\}$. Therefore, by the first isomorphism theorem, $\mrm{PSL}(2,\CC)=\mrm{SL}(2,\CC)/\{\id,-\id\}\cong \mrm{SO}^+(3,1)$.

The Möbius group is well studied with well-known conjugacy classes~\cite{Beardon}. There are five conjugacy classes: identity, elliptic, parabolic, hyperbolic and loxodromic. In the context of the restricted Lorentz group, these correspond to the identity, spatial rotations, null rotations, boosts, and boosts combined with rotations. Without loss of generality, the elliptic conjugacy class is generated by $\mrm{R}_{12}$, the parabolic conjugacy class by $\mrm{NR}_{012}$, the hyperbolic conjugacy class by $\mrm{B}_{01}$, and the loxodromic conjugacy class by $\mrm{B}_{01}\bigoplus\mrm{R}_{23}$.

\subsubsection{Conjugacy classes of the restricted Lorentz group}

\noindent We are now in a position to classify the conjugacy classes of $\mrm{SO}^+(n,1)$. First, we note the conjugacy classes of $\mrm{SO}^+(n,1)$ for $n<3$: $\sop(1,1)$ contains only the identity and hyperbolic classes, whereas $\sop(2,1)$ contains the identity, elliptic, parabolic, and hyperbolic conjugacy classes. Both follow from reducing the available dimensions in the $\sop(3,1)$ case.

For $n>3$, it has been demonstrated that elements of $\sop(n,1)$ are conjugate to one of three canonical forms depending on the eigenvalues of the element~\cite{Abraham1948Lorentz,Cirici2014}. In particular, any element of $\sop(n,1)$ can be reduced to the form $\xi_0\oplus\eta\oplus\id$, where $\xi_0\in\sop(3,1)$ or $\sop(2,1)$, $\eta\in \mrm{SO}(m)$, $\id$ is the $(n-m-3)$ or $(n-m-2)$-dimensional identity matrix respectively, and $\oplus$ is the direct sum of matrices. Furthermore, elements of $\mrm{SO}(m)$ can be reduced to a canonical form~\cite{Cirici2014}: for $m$ even, $\eta=\oplus_{i=1}^{m/2}\eta_i$ and for $m$ odd, $\eta=\oplus_{i=1}^{\lfloor m/2\rfloor}\eta_i\oplus1$, where $\eta_i\in\mrm{SO}(2)$.

\subsubsection{Summary}

\noindent 
\color{black}We write down the non-identity conjugation classes for $n\geq 3$ (cf. classification of the orthochronous components of $\mrm{O}(n,1)$~\cite{Mars:2020ytn,Mars:2020yer}),
\begin{subequations}\label{eqn:: conj classes of classes}
    \begin{align}
        \label{eqn:: elliptic conj}\xi^{k}_E&~=~\begin{cases}
        \mrm{R}_{12} &\;\;\,\, \,k=1\,,\\
            \mrm{R}_{12}\bigoplus_{i=2}^k\mrm{R}_{2i-1\,2i}\ & \;\;\,\, \,k\geq 2\,,
        \end{cases}\\
        \label{eqn:: parabolic conj}\xi^k_P&~=~
        \begin{cases}
            \mrm{NR}_{012} & k=1\,,\\
            \mrm{NR}_{012}\bigoplus_{i=2}^k\mrm{R}_{2i-1\,2i} & k\geq 2\,,
        \end{cases}\\
        \label{eqn:: loxodromic conj}\xi^{l}_L&~=~
        \begin{cases}
            \mrm{B}_{01}&\:l=1\,,\\
            \mrm{B}_{01}\bigoplus_{i=2}^l\mrm{R}_{2i-2\,2i-1} &\: l\geq 2\,,
        \end{cases}
    \end{align}
\end{subequations}for $1\leq k\leq \lfloor\tfrac{n}{2}\rfloor$, $1\leq l\leq\lceil\tfrac{n}{2}\rceil$, where $n+1$ is the spacetime dimension and $\lfloor\cdot\rfloor$ and $\lceil\cdot\rceil$ are the floor and ceiling functions respectively. We will refer to $\xi^k_{\mrm{E}}$ as the elliptic conjugacy class, $\xi^k_{\mrm{P}}$ as the parabolic conjugacy class, and $\xi^l_{\mrm{L}}$ as the loxodromic conjugacy class. It is important to note that the boosts and rotations appearing in~\eqref{eqn:: conj classes of classes} are considered to have non-zero rotation angles, that is to say the scalar coefficients are non-zero. The set of non-identity conjugacy classes of $\sop(n,1)$ are then given by $\{\xi^k_E,\,\xi^k_P,\,\xi^l_L:1\leq k\leq \lfloor\tfrac{n}{2}\rfloor,\,1\leq l\leq \lceil\tfrac{n}{2}\rceil\}$ for $n\geq2$. We remark that, using the terminology of the conjugacy classes of the Möbius group, $\xi_{\mrm{L}}^1$ is the hyperbolic conjugacy class. As a consistency check, we note that this set recovers what we reported earlier for $n=3$. We denote the identity conjugacy class by $\id$.

\subsection{Conjugacy classes of the restricted Poincaré group}\label{subsec:: poincare conj}
We now extend the classification to the restricted Poincaré group.  We represent an element of $\isop(n,1)$ as
\begin{equation}
    g~=~\begin{pmatrix}
        \Lambda&a\\0^\intercal&1
    \end{pmatrix}\,,
\end{equation}where $\Lambda\in\sop(n,1)$, $1\in\RR$, and $a,\,0\in\RR^{n,1}$ and are viewed as column vectors . Then, $\isop(n,1)$ acts on $\RR^{n,1}$ by
\begin{equation}
    g\cdot\xx~=~\begin{pmatrix}
        \Lambda&a\\0^\intercal&1
    \end{pmatrix}\begin{pmatrix}
        \xx\\1
    \end{pmatrix}~=~\begin{pmatrix}
        \Lambda\xx+a\\1
    \end{pmatrix}\,.
\end{equation}
Returning to the generators of the Poincaré Lie algebra, one may perform an infinitesimal transformation directly to the coordinates $\xx^\alpha$ to find a matrix representation. The Lorentz generators ${(M_{\mu\nu})^A}_B$ and translation generators ${(P_\mu)^A}_B$ are given by
\begin{subequations}
    \begin{align}
       {(M_{\mu\nu})^A}_B&~=~\delta^A_\mu\eta_{\nu B}-\delta^A_\nu\eta_{\mu B}\,,\\
       {(P_\mu)^A}_B&~=~\delta^A_\mu\delta^{n+1}_B\,,
    \end{align}
\end{subequations}where $A$, $B=0,\,1,\,\dots,\,n+1$. In this notation, the spacetime rotation generators are given in matrix form by $\mrm{R}_{ij}=M_{ij}$, $\mrm{B}_{0i}=M_{0i}$, $\mrm{NR}_{0ij}=M_{0i}-M_{ij}$.

Consider a linear combination of Killing vectors. This will have a matrix (Lorentz) component $N$ and a vector (translation) component $K$. Under conjugation by $g=(\Lambda,a)$, we have
\begin{equation}\label{eqn:: conjugacy action}
    g\cdot\begin{pmatrix}
        N&K\\0^\intercal&0
    \end{pmatrix}\cdot g^{-1}~=~\begin{pmatrix}
        \Lambda N\Lambda^{-1}&-\Lambda N\Lambda^{-1}a+\Lambda K\\0^\intercal&0
    \end{pmatrix}\,.
\end{equation}
We are now in a position to find the conjugacy classes.

\subsubsection{Temporal translation}\label{subsubsec:: temporal}

\noindent We consider first the timelike translations $\mrm{T}_0=P_0$. One may add a timelike translation to the identity conjugacy class, resulting in the class of inertial trajectories $\mrm{T}_0$. Adding a timelike translation to the loxodromic conjugacy class~\eqref{eqn:: loxodromic conj} results in $\mrm{T}_0\oplus\xi_L^l=\alpha\p_t+(t\p_1+x^1\p_t)+\sum_{i=2}^lb_i(x^{2i-1}\p_{2i-2}-x^{2i-2}\p_{2i-1})$. This linear combination as a matrix results in a matrix part ${N^\mu}_\nu=(\delta^\mu_0\eta_{1\nu}-\delta_1^\mu\eta_{0\nu})+\sum_{i=2}^lb_i(\delta^\mu_{2i-2}\eta_{2i-1\,\nu}-\delta^\mu_{2i-1}\eta_{2i-2\,\nu})$ and a vector part $K^\mu=\alpha \delta_0^\mu$. The loxodromic conjugacy class $\xi_L^l$ contains the same matrix contribution and no vector contribution. We can force the vector part of $\mrm{T}_0\oplus\xi^l_L$, $(-\Lambda N\Lambda^{-1}a+\Lambda K)$, in the conjugation~\eqref{eqn:: conjugacy action} to vanish by choosing $a^\mu=\alpha\eta^{\mu\beta}{(\Lambda^{-1})^1}_\beta$. Therefore, with this choice of $a$,
\begin{align}
    g\cdot\left(\mrm{T}_0\bigoplus\xi^l_L\right)\cdot g^{-1}&~=~(\Lambda,a)\cdot (N,K)\cdot(\Lambda,a)^{-1}\,,\nonumber\\
    &~=~(\Lambda N\Lambda^{-1},0)\,,\nonumber\\
    &~=~(\Lambda,0)\cdot(N,0)\cdot(\Lambda,0)^{-1}\,,\nonumber\\
    &~\sim~\xi^l_L\,.\label{eqn:: T loxo conj}
\end{align}

We consider now the elliptic $\xi^k_E$~\eqref{eqn:: elliptic conj} and parabolic $\xi^k_P$~\eqref{eqn:: parabolic conj} Killing generators. Neither of these can be timelike anywhere. However, both $\mrm{T}_0\oplus\xi^k_E$ and $\mrm{T}_0\oplus\xi^k_P$ can be timelike somewhere. Since conjugation does not change the timelike/null/spacelike nature of a Killing vector, we conclude that $\mrm{T}_0\oplus\xi^k_{E(P)}\not\sim \xi^k_{E(P)}$.

\subsubsection{Spatial translation}\label{subsubsec:: spatial}
\noindent We consider now the spacelike translations $\mrm{S}_m=P_m$ with $1\leq m\leq n$ fixed. A spacelike translation added to the identity conjugacy class results in a spatial curve. Consider now the loxodromic conjugacy class, $\mrm{S}_m\oplus\xi_L^l$. This can be written in terms of Killing vectors as $\mrm{S}_m\oplus\xi^l_L=\alpha\p_m+(x^1\p_t+t\p_t)+\sum_{i=2}^lb_i(x^{2i-1}\p_{2i-2}-x^{2i-2}\p_{2i-1})$. The vector part of the conjugation $(\Lambda,a)\cdot(\mrm{S}_m\oplus\xi^l_L)\cdot(\Lambda,a)^{-1}$ reads
\begin{multline}\label{eqn:: vector space loxo}
    0~=~-{\Lambda^\mu}_0{(\Lambda^{-1})^1}_\beta a^\beta-{\Lambda^\mu}_1{(\Lambda^{-1})^0}_\beta a^\beta+\alpha{\Lambda^
    \mu}_m\\ +\sum_{i=2}^l b_i\left({\Lambda^\mu}_{2i-2}{(\Lambda^{-1})^{2i-1}}_\beta-{\Lambda^\mu}_{2i-1}{(\Lambda^{-1})^{2i-2}}_\beta\right)a^\beta\,.
\end{multline}

In the case $m=1$, this translation is parallel to the boost. By choosing $a^\beta=\alpha\eta^{\beta\rho}{(\Lambda^{-1})^0}_\rho$, one can make the vector contribution~\eqref{eqn:: vector space loxo} vanish. In the case $1<m\leq2l-1$, this translation is parallel to an axis of rotation and can be conjugated away. For $m$ even, $a^\beta=-\alpha/(b_{(m+2)/2})\eta^{\beta\rho}{(\Lambda^{-1})^{m+1}}_\rho$ and for $m$ odd, $a^\beta=\alpha/(b_{(m+1)/2})\eta^{\beta\rho}{(\Lambda^{-1})^{m-1}}_\rho$ will make~\eqref{eqn:: vector space loxo} vanish. However, for $2l-1< m\leq n$  with $l<\lceil\tfrac{n}{2}\rceil$, one is unable to conjugate away $\mrm{S}_m$. If $l=\lceil\tfrac{n}{2}\rceil$, then the result depends on the parity of $n$. If $n$ is odd, then all available spatial dimensions are filled by the boost along $x^1$ and rotations in the remaining $(n-1)/2$ independent planes. By contrast, if $n$ is even, there is then one free axis, parallel to which one may perform a spatial translation.

To summarise, $\mrm{S}_m\oplus\xi_L^l\sim\xi_L^l$ for $1\leq m\leq2l-1$ and $1\leq l<\lceil\tfrac{n}{2}\rceil$. Whereas, for $2l-1< m\leq n$ and $1\leq l\leq\lceil\tfrac{n}{2}\rceil$, $\mrm{S}_m\oplus\xi_L^l$ forms a new conjugacy class.

The analyses for $\mrm{S}_m\oplus\xi_E^k$ and $\mrm{S}_m\oplus\xi_P^k$ are characteristically and computationally similar to the loxodromic case. We summarise the results now. For $1\leq m \leq 2k$ and $1\leq k\leq\lfloor\tfrac{n}{2}\rfloor$, one may choose a translation $a$ suitably to conjugate away the translation $\mrm{S}_m$. However, for $2k<m\leq n$ with $1\leq k\leq \lfloor\tfrac{n}{2}\rfloor$, there is a free axis, parallel to which one may perform a spatial translation, producing two more sets of conjugacy classes, $\mrm{S}_m\oplus\xi_{E(P)}^k$ for $2k<m\leq n$.

We remark that in all cases where one may add a spatial translation, it is parallel to an axis, along which and parallel to which there are no other motions. As such, if one were to add multiple spatial translations, each along axes independent of the other motions, one could choose a $\Lambda$ to align all translations along one axis. This is possible since in each conjugacy class, $\Lambda$ was hitherto arbitrary. Hence, we need only consider one spatial translation.

\subsubsection{Temporal and spatial translation}

\noindent We consider now the case of spatial and temporal translations, $\mrm{T}_0\oplus\xi\bigoplus_i\mrm{S}_i$. If the translations $\mrm{T}_0$ or $\mrm{S}_i$ are parallel to a plane of boost or plane of rotation, they can be conjugated away as seen in Sections~\ref{subsubsec:: temporal} and~\ref{subsubsec:: spatial}. We first consider the loxodromic conjugacy class $\xi=\xi_L^l$~\eqref{eqn:: loxodromic conj}. The timelike translation $\mrm{T}_0$ can be conjugated away so we have $\xi_L^l\bigoplus_i\mrm{S}_i$. Any $\mrm{S}_i$ parallel to the rotations or the boost can be conjugated away. This will either conjugate away all $\mrm{S}_i$ or we are left with $\xi_L^l\bigoplus_{i\geq2l}\mrm{S}_i$. Finally, one can perform rotations in the hyperplanes containing the $\mrm{S}_i$ to align them along one axis, leaving $\xi_L^l\oplus\mrm{S}_{2l}$.

We consider now the parabolic and elliptic conjugacy classes~\eqref{eqn:: elliptic conj} and~\eqref{eqn:: parabolic conj}. Once again, we can conjugate away any translations parallel to a plane of rotation and then rotate in the planes containing the remaining spatial translations, leaving $\mrm{T}_0\oplus\xi_{E(P)}^k\oplus\mrm{S}_{2k+1}$. In this case, we consider the relative magnitudes of the translations $\mrm{T}_0$ and $\mrm{S}_{2k+1}$. Let $\mrm{T}_0\oplus\mrm{S}_{2k+1}=\alpha\p_t-\beta\p_{2k+1}$, then: if $|\alpha|>|\beta|$, this is conjugate to $\mrm{T}_0$; if $|\alpha|<|\beta|$, this is conjugate to $\mrm{S}_{2k+1}$; and if $|\alpha|=|\beta|$, this is a null translation $\mrm{NT}_{0,2k+1}$.

Finally, we consider the identity conjugacy class $\id$. We can perform temporal and spatial translations $\mrm{T}_0\oplus\id\bigoplus_i\mrm{S}_i$. Depending on whether $\mrm{T}_0\bigoplus_i\mrm{S}_i$ is timelike, spacelike, or null, this is conjugate to $\mrm{T}_0$, $\mrm{S}_1$, or $\mrm{NT}_{01}$.

\subsubsection{Summary of conjugacy classes}

\noindent Just as we listed the conjugacy classes of $\sop(n,1)$ for small $n$, we note that $\isop(0,1)$ contains only $\id$ and $\mrm{T}_0$, and $\isop(1,1)$ contains $\id$, $\mrm{T}_0$, $\mrm{B}_{01}$, $\mrm{S}_1$ and $\mrm{NT}_{01}$. We can now list the conjugacy classes of the Poincaré group, $\isop(n,1)$. We first have the conjugacy classes of $\sop(n,1)$,
\begin{equation}
    \{\id,\xi_E^k,\,\xi_P^k,\,\xi_L^l:1\leq k\leq\lfloor\tfrac{n}{2}\rfloor,\,1\leq l\leq\lceil\tfrac{n}{2}\rceil\}\,.
\end{equation}
In addition, we have those with a time translation,
\begin{equation}
    \{\mrm{T}_0,\,\mrm{T}_0\oplus\xi_E^k,\,\mrm{T}_0\oplus\xi_P^k:1\leq k\leq\lfloor\tfrac{n}{2}\rfloor\}\,.
\end{equation}We also have those with a spatial translation, which we may align along the $x^n$\nobreakdash--axis without loss of generality,
\begin{equation}
    \{\mrm{S}_n\,,\xi^k_E\oplus\mrm{S}_n,\,\xi^k_P\oplus\mrm{S}_n,\,\xi_L^l\oplus\mrm{S}_n:1\leq k\leq\lfloor\tfrac{n}{2}\rfloor,\,1\leq l\leq\lceil\tfrac{n}{2}\rceil\}\,,
\end{equation} with the following caveats: $\xi_{E(P)}^k\oplus\mrm{S}_n\sim\xi_{E(P)}^k$ if $n$ is even and $k=\lfloor\tfrac{n}{2}\rfloor$; and $\xi_L^k\oplus\mrm{S}_n\sim\xi_L^k$ if $n$ is odd and $l=\lceil\tfrac{n}{2}\rceil$.
Finally, we have the conjugacy classes with a null translation, which again may be confined to the $\xx^0$\nobreakdash--$\xx^n$ plane without loss of generality,
\begin{equation}
 \{\mrm{NT}_{0n},\,\mrm{NT}_{0n}\oplus\xi_E^k,\,\mrm{NT}_{0n}\oplus\xi_P^k:1\leq k\leq\lfloor\tfrac{n}{2}\rfloor\}\,,
\end{equation} with the caveat that $\mrm{NT}_{0n}\oplus\xi_{E(P)}^k\sim\mrm{T}_0\oplus \xi_{E(P)}^k$ if $n$ is even and $k=\lfloor\tfrac{n}{2}\rfloor$.
\subsection{Stationary trajectories in Minkowski spacetimes}\label{subsec:: stationary traj}
The conjugacy classes of $\isop(n,1)$ which correspond to a stationary trajectory in $\RR^{n,1}$ are those whose associated Killing vector is timelike somewhere. For $n\geq 3$, we list them:
\begin{subequations}\label{eqn:: stationary classification}

    \begin{eqnarray}
        \xi_0&~\equiv~\mrm{T}_0\quad&\textit{inertial motions,}\\
        \xi_\mrm{LM}^l&~\equiv~\xi^l_{L}\quad&\textit{loxodromic motions}~1\leq l\leq\lceil\tfrac{n}{2}\rceil,\\
        \xi_{\mrm{dL}}^l&~\equiv~\xi_{L}^l\oplus \mrm{S}_n\quad&\textit{drifted loxodromic motions}\begin{dcases}
            1\leq l\leq\lceil\tfrac{n}{2}\rceil&n~\text{even},\\
            1\leq l<\lceil\tfrac{n}{2}\rceil&n~\text{odd},
        \end{dcases}\\
        §\xi_{\mrm{SP}}^k&~\equiv~\mrm{T}_0\oplus\xi_P^k\quad&\textit{semicubical parabolic motions,}\\
        \xi_{\mrm{CM}}^k&~\equiv~\mrm{T}_0\oplus\xi_E^k\quad&\textit{circular motions,}  
    \end{eqnarray}
\end{subequations}for $1\leq k \leq\lfloor\tfrac{n}{2}\rfloor$. The names of the conjugacy classes originate from the classification due to Letaw~\cite{Letaw,LetawPfautsch}. We remark two special cases: $\xi^1_{\mrm{LM}}$ is accelerated (Rindler) motion parallel to the $x^1$\nobreakdash--axis and $\xi^1_{\mrm{dL}}$ is drifted Rindler motion~\cite{Good:2020hav}. The semicubical parabolic motions have the spatial projection of a semicubical parabola in the $x^1$\nobreakdash--$x^2$ plane with circular motions in the remaining independent planes. The circular motions exhibit circular motion in each independent plane.

Let $\mrm{TKV}(n)$ denote the set of conjugacy classes of timelike Killing vectors of $\RR^{n,1}$,
\begin{equation}\label{eqn:: TKV set}
    \mrm{TKV}(n)~=~\{\xi_0,\,\xi^l_{\mrm{LM}},\,\xi^l_{\mrm{dL}},\,\xi^k_{\mrm{SP}},\,\xi^k_{\mrm{CM}}:1\leq l\leq\lceil\tfrac{n}{2}\rceil,\,1\leq k\leq\lfloor\tfrac{n}{2}\rfloor\}\,,
\end{equation}then the number of classes of timelike trajectories is given by
\begin{equation}\label{eqn:: TKV size}
    \#\mrm{TKV}(n)~=~1+3\left\lfloor\frac{n}{2}\right\rfloor+\left\lceil\frac{n}{2}\right\rceil\,.
\end{equation}

Considering now the case $n=4$, $\mrm{TKV}(4)=\{\xi_0,\,\xi^1_{\mrm{LM}},\,\xi^2_{\mrm{LM}},\,\xi^1_{\mrm{dL}},\,\xi^2_{\mrm{dL}},\,\xi^1_{\mrm{SP}},\,\xi^2_{\mrm{SP}}\,,\xi^1_{\mrm{CM}}\,,\xi^2_{\mrm{CM}}\}$ with $\#\mrm{TKV}(4)=9$. We will exhibit and classify these trajectories explicitly in the following Section.

\section{Vielbein formulation}\label{sec:: tetrad formulation}
Stationary trajectories can also be defined as the timelike solutions to the Frenet-Serret equations with proper-time-independent curvature invariants. In this Section, we extend the vierbein formalism of Letaw~\cite{Letaw} to a vielbein formulation, applicable to Minkowski spacetime of dimension $n+1$ with $n\geq1$. We present explicitly the stationary trajectories in $\RR^{4,1}$.

\subsection{Frenet-Serret equations in Minkowski spacetimes}\label{sec:: Frenet Serret Equations}
We begin by constructing an orthonormal vielbein $V_a^\mu(\tau)$ for a worldline $\xx^\mu(\tau)$ in $n+1$ Minkowski spacetime, where $\tau$ is the proper time. These are constructed out of derivatives of the worldline with respect to proper time. We assume that the first $n+1$ derivatives are linearly independent and none of the first $n-1$ derivatives are vanishing or null i.e. ${\xx^\mu}^{(k)}(\tau)\,\xx^{(k)}_\mu(\tau)\neq0$ for $k=1,\,\dots,\,n-1$. Orthonormality is imposed by the relation
\begin{equation}\label{eqn:: tetrad orthonormal}
    V_{a\mu}(\tau)V_{b}^\mu(\tau)~=~\eta_{ab}\,.
\end{equation}
The first component of the vielbein is simply the four-velocity, $V_0^\mu(\tau)=\dot{\xx}^\mu(\tau)$. One may construct a family of orthogonal vielbeins $\widetilde{V}_a^\mu(\tau)$ by the Gram-Schmidt process such that
\begin{equation}\label{eqn:: Gram-Schmidt}
    \widetilde{V}_k^\mu(\tau)~=~\xx^{(k+1)\mu}(\tau)-\sum_{j=1}^{k}\frac{\xx^{(k+1)\rho}(\tau)\xx^{(j)}_\rho(\tau)}{\xx^{(j)\sigma}(\tau)\xx^{(j)}_\sigma(\tau)}\xx^{(j)\mu}(\tau)\,.
\end{equation}
The orthonormal vielbeins $V_a^\mu(\tau)$ are then constructed by the normalisation of $\widetilde{V}_a^\mu(\tau)$. The final vielbein is given by
\begin{equation}
    V_{n}^\mu(\tau)~=~\frac{1}{\sqrt{n!}}\varepsilon^{\rho_0\rho_1\dots\rho_{n-1}\mu}V_{0\rho_0}V_{1\rho_1}\dots V_{n-1\,\rho_{n-1}}\,.
\end{equation}
From now on, we suppress the dependence on the proper time.

Differentiation of the orthonormality condition~\eqref{eqn:: tetrad orthonormal} yields
\begin{equation}\label{eqn:: orthonormal diff}
    \dot{V}_{a\mu}V_b^\mu+V_{a\mu}\dot{V}_b^\mu~=~0.
\end{equation} Since the vielbeins form a basis, one may write the proper time derivatives in the basis of vielbeins,
\begin{equation}\label{eqn:: orthodot}
    \dot{V}_{a}^\mu~=~{K_a}^b(\tau)V_b^\mu\,.
\end{equation}

Combining equations~\eqref{eqn:: orthonormal diff} and~\eqref{eqn:: orthodot} informs us that the matrix $K_{ab}$ is antisymmetric. Furthermore, since each $V_{b}^\mu$ is constructed out of the first $b+1$ derivatives of the worldline, whereas $\dot{V}_a^\mu$ is constructed out of the first $a+2$ derivatives, we have the ${K_a}^b$ vanishes for $b>a+1$. This tells us that the only non-vanishing components are the off-diagonal components and one may write this matrix as
\begin{equation}\label{eqn:: invariant matrix}
    K_{ab}(\tau)~=~\chi_a(\tau)\delta_{a,b-1}-\chi_b(\tau)\delta_{b,a-1}\,,
\end{equation}where $\delta_{ab}$ is the Kronecker delta. 

We will consider only the case where $\chi_a$ are constant in $\tau$ and when $V_0^\mu$ is future-directed. These $\chi_a$ are then referred to as the curvature invariants. Combining~\eqref{eqn:: orthodot} and~\eqref{eqn:: invariant matrix} then yields a set of equations referred to as the Frenet-Serret equations.

This explicit form of the matrix of curvature invariants~\eqref{eqn:: invariant matrix} enables us to rewrite the Frenet-Serret equations as
\begin{subequations}\label{eqn:: frenet-serret}
    \begin{align}
        V_1^\mu&~=~\frac{1}{\chi_0}\dot{V}_0^\mu\,,\\
        V_2^\mu&~=~\frac{1}{\chi_0\chi_1}\left(\ddot{V}_0^\mu-\chi_0^2V_0^\mu\right)\,,\\
        V_a^\mu&~=~\frac{1}{\chi_{a-1}}\left(\dot{V}_{a-1}^\mu+\chi_{a-2}V_{a-2}^\mu\right)\,,\quad a=3,\,\dots,\,n\,,\\
        \dot{V}^\mu_n&~=~-\chi_{n-1}V_{n-1}^\mu\,.\label{eqn:: final coupled ODE}
    \end{align}
\end{subequations}Note that setting any $\chi_a=0$ renders the Frenet-Serret equations ill defined. We discuss this further in Appendix~\ref{app:: generalised eqn}. The Frenet-Serret equations~\eqref{eqn:: frenet-serret} enable one to write each vielbein as an ordinary differential equation for $V_0^\mu$. One sees that these differential equations can be written down explicitly in the general case,
\begin{equation}\label{eqn:: general ODE}
    V_a^\mu~=~\frac{1}{\prod_{i=0}^{a-1}\chi_i}\sum_{q=0}^{\left\lfloor \tfrac{a}{2}\right\rfloor} b_{2q}^a\frac{\dd^{a-2q}}{\dd \tau^{a-2q}}V_0^{\mu}\,,\quad a=3,\,\dots,\,n\,,
\end{equation} where the coefficients $b_{2q}^a$ are defined as follows,
\begin{subequations}\label{eqn:: b coeff}
    \begin{align}
        b_0^a&~=~1\,,\\
        b_2^a&~=~\sum_{i,j=0}^{a-2}\eta_{ij}\chi_i\chi_j\,,\\
        b_{2q}^a&~=~\sum_{p_1=2q-2}^{a-2}\chi_{p_1}^2\sum_{p_2=2q-4}^{p_1-2}\chi_{p_2}^2\dots\sum_{i,j=0}^{p_{q-1}-2}\eta_{ij}\chi_i\chi_j\,.\label{eqn:: b general coeff}
    \end{align}
\end{subequations} The dots in~\eqref{eqn:: b general coeff} represent successive insertions of terms of the form $\sum_{p_k=2q-2k}^{p_{k-1}-2}\chi_{p_k}^2$. For example, $b^a_{8}=\sum_{p_1=6}^{a-2}\chi_{p_1}^2\sum_{p_2=4}^{p_1-2}\chi_{p_2}^2\sum_{p_3=2}^{p_2-2}\chi_{p_3}^2\sum_{i,j=0}^{p_3-2}\eta_{ij}\chi_i\chi_j$. One may prove~\eqref{eqn:: general ODE} using strong induction and using the relation $b_{2q}^m+\chi^2_{m-1}b^{m-1}_{2(q-1)}=b_{2q}^{m+1}$, which one may derive by expanding~\eqref{eqn:: b coeff}.

In Appendix~\ref{app:: generalised eqn}, we simplify the Frenet-Serret equations and find the ordinary differential equation satisfied by $V^\mu_0$. The generalised Frenet-Serret equations in $n+1$ Minkowski spacetime are therefore given by
\begin{subequations}\label{eqn:: simp frenet-serret}
    \begin{align}
        V_1^\mu&~=~\frac{1}{\chi_0}\dot{V}_0^\mu\,,\\
        V_2^{\mu}&~=~\frac{1}{\chi_0\chi_1}\left(\ddot{V}_0^\mu-\chi_0^2V_0^\mu\right)\,,\\
        V_a^\mu&~=~\frac{1}{\chi_{a-1}}\left(\dot{V}_{a-1}^\mu+\chi_{a-2}V_{a-2}^\mu\right)\,,\quad a=3,\,\dots,\, n\,,\\
        \label{eqn:: V0 ODE}V_{n+1}^\mu&~=~0\,,
    \end{align}
\end{subequations} with $V_a^\mu$ given by~\eqref{eqn:: general ODE} for $a=3,\,\dots,\,n+1$. Using terminology from differential equations, the characteristic equation of~\eqref{eqn:: V0 ODE} is then
\begin{equation}\label{eqn:: characteristic eqn}
    \sum_{q=0}^{\left\lfloor\tfrac{n+1}{2}\right\rfloor}b_{2q}^{n+1}m^{n+1-2q}~=~0\,,
\end{equation}which has definite parity. 
The coefficients of the general solution can then be fixed by an initial condition, for which we may adopt
\begin{equation}\label{eqn:: initial conditions}
   \left.V_a^\mu(\tau)\right|_{\tau=0}~=~\delta^\mu_a\,.
\end{equation} One may remark that under a Poincaré transformation of the worldline $\xx^\mu\mapsto\xx'^\mu={\Lambda^\mu}_\nu\xx^\nu+b^\nu$, the vielbeins transform as $V_a^\mu\mapsto {V'}_a^\mu={\Lambda^\mu}_\nu V_a^\nu$. These transformed vielbeins are also an orthonormal basis, obeying the orthonormality condition~\eqref{eqn:: tetrad orthonormal} ${V'}_{a\mu}{V'}^\mu_b=\eta_{ab}$. The choice of the direction of the tangent vector at $\tau=0$~\eqref{eqn:: initial conditions} determines which orthonormal basis one uses. This is the geometric equivalent of the conjugacy class of a Killing vector, as found in Section~\ref{subsec:: stationary traj}.

\textit{A priori}, one is able to express solutions to~\eqref{eqn:: characteristic eqn} in terms of radicals only for $n\leq 8$~\cite{Galois}. We demonstrate this extended formalism in the following Section to explicitly calculate and classify the stationary trajectories in $4+1$ Minkowski spacetime.

\subsection{Example: $4+1$ Minkowski spacetime}\label{sec:: 4+1 examples}
In this Section, we use the formalism of Section~\ref{sec:: Frenet Serret Equations} to present the stationary trajectories in $4+1$ Minkowski spacetime. We demonstrate the equivalence between solutions to the Frenet-Serret equations with constant curvature invariants and the integral curves of timelike Killing vectors in $4+1$ dimensions. We classify the resulting trajectories into nine equivalence classes.

The characteristic equation~\eqref{eqn:: characteristic eqn} in $4+1$ dimensions reads
\begin{align}\label{eqn:: 4 1 char eqn}
    0~=~m(m^4-2am^2-b)\,,
\end{align} where $2a=-b_2^5$ and $b=-b_4^5$, each given by~\eqref{eqn:: b general coeff}. This has the solutions $m=0$ and $m^2=a\pm\sqrt{a^2+b}$, which we write as $m^2=\sqrt{a^2+b}+a$ or $m^2=-(\sqrt{a^2+b}-a)$. Hence $m=0,~\pm R_+,~\pm\ii R_-$, where $R^2_\pm=\sqrt{a^2+b}\pm a$. The general solution for $V_0^\mu$ is then
\begin{equation}\label{eqn:: 4+1 V0}
    V_0^\mu~=~A^\mu+B^\mu\cosh(R_+\tau)+C^\mu\sinh(R_+\tau)+D^\mu\cos(R_-\tau)+E^\mu\sin(R_-\tau)\,.
\end{equation}
Using the initial conditions~\eqref{eqn:: initial conditions}, the coefficients are found to be
\begin{subequations}\label{eqn:: general 4+1 coeffs}
\begin{align}
    A^\mu&~=~\left(1-\frac{\chi_0^2}{b}(\chi_2^3+\chi_3^2),0,-\frac{\chi_0\chi_1\chi_3^2}{b},0,-\frac{\chi_0\chi_1\chi_2\chi_3}{b}\right)\,,\\
    B^\mu&~=~\frac{1}{R^2}\left(\frac{\chi_0^2}{R_+^2}(\chi_0^2-\chi_1^2+R_-^2),0,\frac{\chi_0\chi_1}{R_+^2}(\chi_0^2-\chi_1^2-\chi_2^2+R_-^2),0,\frac{\chi_0\chi_1\chi_2\chi_3}{R_+^2}\right)\,,\\
    C^\mu&~=~\frac{1}{R^2}\left(0,\frac{\chi_0}{R_+}(\chi_0^2-\chi_1^2+R_-^2),0,\frac{\chi_0\chi_1\chi_2}{R_+},0\right)\,,\\
    D^\mu&~=~\frac{1}{R^2}\left(\frac{\chi_0^2}{R_-^2}(\chi_0^2-\chi_1^2-R_+^2),0,\frac{\chi_0\chi_1}{R_-^2}(\chi_0^2-\chi_1^2-\chi_2^2-R_+^2),0,\frac{\chi_0\chi_1\chi_2\chi_3}{R_-^2}\right)\,,\\
    E^\mu&~=~\frac{1}{R^2}\left(0,-\frac{\chi_0}{R_-}(\chi_0^2-\chi_1^2-R_+^2),0,-\frac{\chi_0\chi_1\chi_2}{R_-},0\right)\,,
\end{align}
\end{subequations} where $R^2=R_+^2+R_-^2$.

We explicitly calculate the stationary trajectories in $4+1$ Minkowski spacetime in Appendix~\ref{app:: Motions and Lorentz}. We report the results case-by-case. For $m\leq n$, the stationary trajectories of $\RR^{m,1}$ are embedded in $\RR^{n,1}$. Hence to find all stationary trajectories in $\RR^{n,1}$, one solves the Frenet-Serret equations~\eqref{eqn:: simp frenet-serret} for each $m=0,~1,~\dots,~n$.

The stationary trajectories of $4+1$ Minkowski spacetime are as follows: Case 0: the class of inertial trajectories~\eqref{eqn:: 0}. Case I: the class of Rindler trajectories~\eqref{eqn:: I}. Case IIa: the class of drifted Rindler motions~\eqref{eqn:: IIa}. Case IIb: the class of motions with semicubical parabolic spatial projection~\eqref{eqn:: IIb}. Case IIc: the class of circular motions~\eqref{eqn:: IIc}. Case III: the class of loxodromic motions~\eqref{eqn:: III}. Case IVa: the class of drifted loxodromic motions in the $x^1$\nobreakdash--$x^2$ plane with circular motion in the $x^3$\nobreakdash--$x^4$ plane~\eqref{eqn:: IVa drifted rindler}. Case IVb: the class of motions with semicubical parabolic spatial projection in the $x^1$\nobreakdash--$x^2$ plane with circular motion in the $x^3$\nobreakdash--$x^4$ plane~\eqref{eqn:: IVa semicubical}. Case IVc: the class of circular motions in the $x^1$\nobreakdash--$x^2$ plane with circular motion in the $x^3$\nobreakdash--$x^4$ plane~\eqref{eqn:: IVa circular}.

As expected from the previous calculation of $\#\mrm{TKV}(4)$ in Section~\ref{subsec:: stationary traj}, there are nine classes of stationary trajectory. In agreement with~\cite{Letaw,LetawPfautsch}, we recover the six classes of stationary trajectory in $3+1$ Minkowski spacetime.

Since stationary trajectories of Minkowski spacetime are both timelike solutions to the Frenet-Serret equations and the integral curves of timelike Killing vectors, we finish by unifying the two frameworks and present the stationary trajectories with their respective timelike Killing vector according to the classification~\eqref{eqn:: stationary classification},
\begin{subequations}
\begin{flalign}
    &\text{Case $0$: ~~\,Inertial~\eqref{eqn:: 0}}\,\xi_0~=~\mrm{T}_0\,,\\
    &\text{Case I: \,\,\,~~Rindler~\eqref{eqn:: I}}\,\xi_{\mrm{RM}}^1~=~\mrm{B}_{01}\,,\\
    &\text{Case IIa: \,Drifted Rindler~\eqref{eqn:: IIa}}\, \xi_{\mrm{dR}}^1~=~\mrm{B}_{01}\bigoplus\mrm{S}_2\,,\\
    &\text{Case IIb: \,Semicubical parabolic~\eqref{eqn:: IIb}}\,\xi^1_{\mrm{SP}}~=~\mrm{T}_0\bigoplus \mrm{NR}_{012}\,,\\
    &\text{Case IIc: \,Circular~\eqref{eqn:: IIc}}\,\xi_{\mrm{CM}}^1~=~\mrm{T}_0\bigoplus\mrm{R}_{12}\,,\\
    &\text{Case III: \,~Loxodromic~\eqref{eqn:: III}}\,\xi_{\mrm{RM}}^2~=~\mrm{B}_{01}\bigoplus \mrm{R}_{23}\,,\\
    &\text{Case IVa: Drifted loxodromic with circular~\eqref{eqn:: IVa drifted rindler}}\,\xi_{\mrm{dR}}^2~=~\mrm{B}_{01}\bigoplus\mrm{S}_2\bigoplus\mrm{R}_{34}\,,\\
    &\text{Case IVb: Semicubical parabolic with circular~\eqref{eqn:: IVa semicubical}}\,\xi_{\mrm{SP}}^2~=~\mrm{T}_0\bigoplus\mrm{NR}_{012}\bigoplus\mrm{R}_{34}\,,\\
    &\text{Case IVc: \,Double circular motion~\eqref{eqn:: IVa circular}}\,\xi_{\mrm{CM}}^2=~\mrm{T}_0\bigoplus\mrm{R}_{12}\bigoplus\mrm{R}_{34}\,.
\end{flalign}
\end{subequations}

\section{Conclusions}
In this paper, we determined the conjugacy classes of the restricted Poincaré group $\isop(n,1)$ and used this to classify the stationary trajectories in Minkowski spacetimes. We found there were five classes of trajectories, which we name: the inertial motions, the loxodromic motions, the drifted loxodromic motions, the semicubical parabolic motions, and the circular motions. Each type of trajectory has conjugacy classes with qualitatively similar motion. The Rindler and drifted Rindler motions are special cases of the loxodromic and drifted loxodromic motions. We then generalised the work of Frenet~\cite{Frenet1852}, Serret~\cite{Serret1851}, Jordan~\cite{Jordan}, and Letaw~\cite{Letaw} to provide a framework for the computation of stationary trajectories in terms of their curvature invariants in Minkowski spacetime. In doing so, we have provided the ordinary differential equation satisfied by the four-velocity of the stationary worldline. We finally utilised this framework to present explicitly the stationary trajectories in $4+1$ Minkowski spacetime.

Minkowski spacetime is a space of zero curvature. A natural extension of this work would be to the spacetimes of constant positive or negative curvature, de Sitter and anti-de Sitter spacetimes respectively~\cite{HawkingEllis}. Previous work in this direction includes the study of the Frenet-Serret equations in general curved spacetimes~\cite{K:2023sex}. The isometry group of $n+1$ de Sitter spacetime is the de Sitter group, whose connected component is isomorphic to $\mrm{SO}^+(n+1,1)$. The classification of the stationary vacuum states in de Sitter~\cite{Parikh:2012ny}, as well as the classification of the conjugacy classes of the restricted Lorentz group in Section~\ref{subsec:: lorentz conj} would therefore be relevant and adaptable to this classification. However, the connected component of the isometry group of $n+1$ anti-de Sitter spacetime, the anti-de Sitter group, is isomorphic to $\mrm{SO}^+(n,2)$. This change in signature in comparison to the Minkowski or de Sitter isometry groups means that new techniques will be required in classifying the conjugacy classes of anti de Sitter spacetime.

\begin{acknowledgments}
    CRDB is indebted to Leo Parry and Jorma Louko for invaluable discussions. I thank Carlos Peón-Nieto and Marc Mars for bringing the work in~\cite{BURGOYNE1977339,cushman2006adjoint,Mars:2020ytn,Mars:2020yer} to my attention, Prasant Samantray for bringing the work in~\cite{Parikh:2012ny,Parikh:2011aa} to my attention, and Hari K for bringing the work in~\cite{K:2023sex} to my attention. I thank the anonymous referees for helpful comments and for bringing the work in~\cite{townsend_stationary} to my attention. For the purpose of open access, the authors have applied a CC BY public copyright licence to any Author Accepted Manuscript version arising.
\end{acknowledgments}

\appendix
\section{Generalised Frenet-Serret equations}\label{app:: generalised eqn}
In this Section, we simplify the generalised Frenet-Serret equations~\eqref{eqn:: frenet-serret}. We begin with the equations~\eqref{eqn:: final coupled ODE} and the general ordinary differential equation defining $V^\mu_a$ in terms of $V^\mu_0$~\eqref{eqn:: general ODE}. We insert~\eqref{eqn:: general ODE} into~\eqref{eqn:: final coupled ODE}, resulting in an ordinary differential equation for $V_0^\mu$,

\begin{equation}
   0~=~\frac{1}{\prod_{i=0}^{n-1}\chi_i}\left(\sum_{q=0}^{\left\lfloor\tfrac{n}{2}\right\rfloor}b_{2q}^n\frac{\dd^{n+1-2q}}{\dd\tau^{n+1-2q}}V_0^\mu+\sum_{q=0}^{\left\lfloor\tfrac{n-1}{2}\right\rfloor}\chi^2_{n-1}b_{2q}^{n-1}\frac{\dd^{n-1-2q}}{\dd\tau^{n-1-2q}}V_0^\mu\right)\,.
\end{equation}This can brought into a more familiar form,
\begin{align}
    0~&=~\frac{1}{\prod_{i=0}^{n-1}\chi_i}\left(\sum_{q=0}^{\left\lfloor\tfrac{n}{2}\right\rfloor}b_{2q}^n\frac{\dd^{n+1-2q}}{\dd\tau^{n+1-2q}}V_0^\mu+\sum_{q=0}^{\left\lfloor\tfrac{n-1}{2}\right\rfloor}\chi^2_{n-1}b_{2q}^{n-1}\frac{\dd^{n-1-2q}}{\dd\tau^{n-1-2q}}V_0^\mu\right)\,,\nonumber\\
    &\overset{\textbf{(a)}}{=}~\frac{1}{\prod_{i=0}^{n-1}\chi_i}\left(\sum_{q=0}^{\left\lfloor\tfrac{n}{2}\right\rfloor}b_{2q}^n\frac{\dd^{n+1-2q}}{\dd\tau^{n+1-2q}}V_0^\mu+\sum_{q=1}^{\left\lfloor\tfrac{n+1}{2}\right\rfloor}\chi^2_{n-1}b_{2(q-1)}^{n-1}\frac{\dd^{n+1-2q}}{\dd\tau^{n+1-2q}}V_0^\mu\right)\,,\nonumber\\
    &\overset{\textbf{(b)}}{=}~\frac{1}{\prod_{i=0}^{n-1}\chi_i}\left(\frac{\dd^{n+1}}{\dd\tau^{n+1}}V_0^\mu+\sum_{q=1}^{\left\lfloor\tfrac{n+1}{2}\right\rfloor}\left[b_{2q}^n+\chi_{n-1}^2b_{2(q-1)}^{n-1}\right]\frac{\dd^{n+1-2q}}{\dd\tau^{n+1-2q}}V_0^\mu\right)\,,\nonumber\\
    &\overset{\textbf{(c)}}{=}~\frac{\chi_n}{\prod_{i=0}^n\chi_i}\sum_{q=0}^{\left\lfloor\tfrac{n+1}{2}\right\rfloor}b^{n+1}_{2q}\frac{\dd^{n+1-2q}}{\dd\tau^{n+1-2q}}V_0^\mu\,,\nonumber\\
    &\overset{\textbf{(d)}}{=}\chi_nV_{n+1}^\mu\,.    
\end{align} In equality \textbf{(a)}, we changed summation variable $q\mapsto q+1$ in the second summation. In equality \textbf{(b)}, we isolated the $q=0$ term and used that for odd integers $\lfloor\tfrac{n}{2}\rfloor=\lfloor\tfrac{n+1}{2}\rfloor$. For odd integers, consider the effect of replacing $\lfloor\tfrac{n}{2}\rfloor$ by $\lfloor\tfrac{n+1}{2}\rfloor$. Let $n=2m-1$, an odd integer. Then $\lfloor\tfrac{n+1}{2}\rfloor=m$. Hence, the coefficient of this term would be $b_{2m}^{2m-1}$. Calculating this using~\eqref{eqn:: b general coeff}, one finds that the first summation is over $\sum_{2m-2}^{2m-3}$, which identically vanishes. Hence, one may replace $\lfloor\tfrac{n}{2}\rfloor$ by $\lfloor\tfrac{n+1}{2}\rfloor$ in the first sum. In equality \textbf{(c)}, we used the relation $b_{2q}^m+\chi_{m-1}^2b_{2(q-1)}^{m-1}=b_{2q}^{m+1}$ and then combined the two terms under one summation and finally multiplied by $\chi_n/\chi_n$. In equality \textbf{(d)}, we recognised that the resulting summation was~\eqref{eqn:: general ODE} in the case $a=n+1$. 

Therefore, one may find the ordinary differential equation for $V_0^\mu$ (and hence the four-velocity) by forcing $V_{n+1}^\mu$ to vanish. This fully determines the Frenet-Serret equations.

\section{Stationary trajectories in $4+1$ Minkowski spacetime}\label{app:: Motions and Lorentz}
In this Section, we present the stationary trajectories in $4+1$ Minkowski spacetime. One first solves the ordinary differential equation for $V_0^\mu$, then calculates the constants of integration using~\eqref{eqn:: initial conditions} and finally brings the motion into a more familiar form by a suitable Lorentz transformation.

If one sets $\chi_a=0$, then the Frenet-Serret equations~\eqref{eqn:: simp frenet-serret} are no longer well defined. The geometric effect of setting $\chi_a=0$ is to confine the motion to $\RR^{a+1,1}$. Note, however, that the stationary trajectories of $\RR^{m,1}$ are also present in $\RR^{n,1}$ for $m\leq n$. Therefore, to calculate the stationary trajectories present in $\RR^{n,1}$, one solves the Frenet-Serret equations~\eqref{eqn:: simp frenet-serret} for each $m\leq n$. The trajectories are then written via the inclusion map,
\begin{subequations}
    \begin{align}
        \RR^{m,1}&\hookrightarrow\RR^{n,1}\,,\\
        (\xx^0,\,\dots,\,\xx^m)&\mapsto(\xx^0,\,\dots,\,\xx^m,\,0,\,\dots,\,0)\,,
    \end{align}
\end{subequations}where $(0,\,\dots,\,0)$ represents ($n-m$)\nobreakdash--many zeros (in the case $m=n$, then there are no zeros present).

We present the stationary trajectory in cases. \textbf{Case $m$} gives the solution(s) to the Frenet-Serret equations~\eqref{eqn:: simp frenet-serret} in $\RR^{m,1}$.

\setlength{\parindent}{0cm}
\paragraph*{\textbf{Case $0$}~---}The class of inertial trajectories,
\begin{equation}\label{eqn:: 0}
        V_0^\mu~=~\dot{\xx}^\mu=(1,0,0,0,0)\,.
    \end{equation}
    
\paragraph*{\textbf{Case I}~---}$\chi_0>0$. Rindler motion,
\begin{equation}\label{eqn:: I}
    V_0^\mu~=~(\cosh(\chi_0\tau),\sinh(\chi_0\tau),0,0,0)\,.
\end{equation}

\paragraph*{\textbf{Case II}~---}The solutions to the Frenet-Serret equations in $2+1$ have two free parameters, the curvature invariants $\chi_0$ and $\chi_1$. The classification of the stationary trajectory depends on their relation.
\paragraph*{\textbf{Case IIa}~---}$\chi_0>|\chi_1|>0$. After a suitable Lorentz transformation, this is drifted Rindler motion~\cite{Good:2020hav},
    \begin{equation}
        \label{eqn:: IIa}
        V_0^\mu~=~\frac{1}{\sqrt{\chi_0^2-\chi_1^2}}\left(\chi_0\cosh(\sqrt{\chi_0^2-\chi_1^2}\,\tau),\chi_0\sinh(\sqrt{\chi_0^2-\chi_1^2}\,\tau),\chi_1,0,0\right)\,.
    \end{equation}
    
\paragraph*{\textbf{Case IIb}~---}$|\chi_0|=|\chi_1|\neq0$,
\begin{equation}\label{eqn:: IIb}
    V_0^\mu~=~\left(1+\tfrac{1}{2}\chi_0^2\tau^2,\chi_0\tau,\tfrac{1}{2}\chi_0^2\tau^2,0,0\right)\,,
\end{equation}whose spatial profile is that of the semicubical parabola $y^2=\tfrac{2}{9}\chi_0x^3$.

\paragraph*{\textbf{Case IIc}~---}$\chi_1>|\chi_0|>0$. After a suitable Lorentz transformation, this is circular motion in the $x^1$\nobreakdash--$x^2$ plane.
\begin{equation}
 \label{eqn:: IIc}
        V_0^\mu~=~\frac{1}{\sqrt{\chi_1^2-\chi_0^2}}\left(\chi_1,-\chi_0\sin(\sqrt{\chi_1^2-\chi_0^2}\,\tau),\chi_0\cos(\sqrt{\chi_1^2-\chi_0^2}\,\tau),0,0\right)\,.   
\end{equation}
\newline
In the following cases, we give the Lorentz transformation explicitly owing to their more involved calculations.
\paragraph*{\textbf{Case III}~---}The general solution to~\eqref{eqn:: V0 ODE} is
\begin{subequations}
    \begin{align}
    V_0^\mu&~=~B^\mu\cosh(R_+\tau)+C^\mu\sinh(R_+\tau)+D^\mu\cos(R_-\tau)+E^\mu\sin(R_-\tau)\,,\\
        B^\mu&~=~\frac{1}{R^2}\left(R_-^2+\chi_0^2,0,\chi_0\chi_1,0,0\right)\,,\\
        C^\mu&~=~\frac{1}{R^2}\left(0,\frac{\chi_0}{R_+}(\chi_0^2-\chi_1^2+R_-^2),0,\frac{\chi_0\chi_1\chi_2}{R_+},0\right)\,,\\
        D^\mu&~=~\frac{1}{R^2}\left(R_+^2-\chi_0^2,0,-\chi_0\chi_1,0,0\right)\,,\\
        E^\mu&~=~\frac{1}{R^2}\left(0,-\frac{\chi_0}{R_-}(\chi_0^2-\chi_1^2-R_+^2),0,-\frac{\chi_0\chi_1\chi_2}{R_-},0\right)\,.
    \end{align}
\end{subequations}
This is loxodromic motion, which may more clearly be seen by the following Lorentz transformation,
\begin{subequations}
\begin{align}
    {\Lambda^\mu}_\nu&~=~\begin{pmatrix}
        \alpha&0&\beta&0&0\\
        0&\gamma&0&\delta&0\\
        0&C&0&D&0\\
        A&0&B&0&0\\
        0&0&0&0&1
    \end{pmatrix}\,,\\
    \alpha&~=~\frac{\Delta}{R}\,,\quad
    \beta~=~\frac{\Delta(R_+^2-\chi_0^2)}{\chi_0\chi_1R}\,,\quad
    \gamma~=~\frac{\Delta R_+}{\chi_0 R}\,,
    \delta~=~-\frac{\Delta R_+(\chi_0^2-\chi_1^2-R_+^2)}{\chi_0\chi_1\chi_2R}\,,\nonumber\\
    A&~=~\frac{\chi_0\chi_1}{\Delta R}\,,\quad
    B~=~-\frac{\Delta}{R}\,,\quad
    C~=~-\frac{\chi_1R_-}{\Delta R}\,,\quad
    D~=~\frac{R_-(\chi_0^2-\chi_1^2+R_-^2)}{\chi_2\Delta R}\,,\quad
    \Delta^2~=~R_-^2+\chi_0^2\,,
    \end{align}
\end{subequations}
    \begin{equation}
       \label{eqn:: III}
    {\Lambda^\mu}_\nu V_0^\nu~=~\frac{1}{R}\left(\Delta\cosh(R_+\tau),\Delta\sinh(R_+\tau),-\frac{\chi_0\chi_1}{\Delta}\sin(R_-\tau),\frac{\chi_0\chi_1}{\Delta}\cos(R_-\tau),0\right)\,. 
    \end{equation}
\paragraph*{\textbf{Case IV}~---} The classification of the solutions to the Frenet-Serret equations in $4+1$ Minkowski spacetime depends on the relationship between the four curvature invariants. In particular, on the sign of $b$~\eqref{eqn:: 4 1 char eqn}.
\paragraph*{\textbf{Case IVa}~---}$b=\chi_2^2\chi_0^2+\chi_3^2(\chi_0^2-\chi_1^2)>0$. The four-velocity is given by
\begin{equation}\label{eqn:: Case Va}
    V_0^\mu~=~A^\mu+B^\mu\cosh(R_+\tau)+C^\mu\sinh(R_+\tau)+D^\mu\cos(R_-\tau)+E^\mu\sin(R_-\tau)\,,
\end{equation}
where
\begin{subequations}
\label{eqn:: Case Va coefficients}
\begin{align}
    A^\mu&~=~\left(1-\frac{\chi_0^2}{b}(\chi_2^3+\chi_3^2),0,-\frac{\chi_0\chi_1\chi_3^2}{b},0,-\frac{\chi_0\chi_1\chi_2\chi_3}{b}\right)\,,\\
    B^\mu&~=~\frac{1}{R^2}\left(\frac{\chi_0^2}{R_+^2}(\chi_0^2-\chi_1^2+R_-^2),0,\frac{\chi_0\chi_1}{R_+^2}(\chi_0^2-\chi_1^2-\chi_2^2+R_-^2),0,\frac{\chi_0\chi_1\chi_2\chi_3}{R_+^2}\right)\,,\\
    C^\mu&~=~\frac{1}{R^2}\left(0,\frac{\chi_0}{R_+}(\chi_0^2-\chi_1^2+R_-^2),0,\frac{\chi_0\chi_1\chi_2}{R_+},0\right)\,,\\
    D^\mu&~=~\frac{1}{R^2}\left(\frac{\chi_0^2}{R_-^2}(\chi_0^2-\chi_1^2-R_+^2),0,\frac{\chi_0\chi_1}{R_-^2}(\chi_0^2-\chi_1^2-\chi_2^2-R_+^2),0,\frac{\chi_0\chi_1\chi_2\chi_3}{R_-^2}\right)\,,\\
    E^\mu&~=~\frac{1}{R^2}\left(0,-\frac{\chi_0}{R_-}(\chi_0^2-\chi_1^2-R_+^2),0,-\frac{\chi_0\chi_1\chi_2}{R_-},0\right)\,.
\end{align}
\end{subequations}
Given the classifications in Section~\ref{subsec:: stationary traj}, one may hope to identify this motion as a boost, combined with a drift and circular motion. We make an ansatz of the desired form of the four-velocity and find the appropriate Lorentz transformation,
\begin{equation}\label{eqn:: defining Lorentz}
    V_0^\mu~=~{\Lambda^\mu}_\nu\widetilde{V}^\nu_0~=~\begin{pmatrix}
        \tfrac{B^0}{\alpha}&0&\tfrac{A^0}{\beta}&0&\tfrac{D^0}{\gamma}\\
        0&\tfrac{C^1}{\alpha}&0&-\tfrac{E^1}{\gamma}&0\\
        \tfrac{B^2}{\alpha}&0&\tfrac{A^2}{\beta}&0&\tfrac{D^2}{\gamma}\\
        0&\tfrac{C^3}{\alpha}&0&-\tfrac{E^3}{\gamma}&0\\
        \tfrac{B^4}{\alpha}&0&\tfrac{A^4}{\beta}&0&\tfrac{D^4}{\gamma}
    \end{pmatrix}\begin{pmatrix}
        \alpha\cosh(R_+\tau)\\
        \alpha\sinh(R_+\tau)\\
        \beta\\
        -\gamma\sin(R_-\tau)\\
        \gamma\cos(R_-\tau)
    \end{pmatrix}\,.
\end{equation}
By imposing that $\Lambda$ is a Lorentz transformation, one may identify the coefficients $\alpha$, $\beta$, and $\gamma$ as
\begin{subequations}
    \begin{align}
        \alpha&~=~\sqrt{-B^\mu B_\mu}~=~\sqrt{C^\mu C_\mu}\,,\\
        \beta&~=~\sqrt{A^\mu A_\mu}\,,\\
        \gamma&~=~\sqrt{D^\mu D_\mu}~=~\sqrt{E^\mu E_\mu}\,.
    \end{align}
\end{subequations} Intermediate steps include the verification that $A^\mu D_\mu=C^\mu E_\mu=A^\mu B_\mu = B^\mu D_\mu=0$. This brings the four-velocity into the more recognisable form
\begin{multline}\label{eqn:: IVa drifted rindler}
    V_0^\mu~=~\left(\sqrt{-B^\mu B_\mu}\cosh(R_+\tau),\sqrt{-B^\mu B_\mu}\sinh(R_+\tau),\right.\\\left.\sqrt{A^\mu A_\mu}, -\sqrt{D^\mu D_\mu}\sin(R_-\tau),\sqrt{D^\mu D_\mu}\cos(R_-\tau)\right)\,,
\end{multline}
corresponding to a boost along the $x^1$\nobreakdash--axis, a drift in the $x^2$\nobreakdash--axis and circular motion in the $x^3$\nobreakdash--$x^4$ plane.

\paragraph*{\textbf{Case IVb}~---}$b=0\iff \chi_0^2=\tfrac{\chi_1\chi_3}{\chi_2^2+\chi_3^2}$. In this case, one finds $2a=\chi_0^2-\chi_1^2-\chi_2^2-\chi_3^2=-\tfrac{(\chi_2^2+\chi_3^2)^2+\chi_1^2\chi_2^2}{\chi_2^2+\chi_3^2}<0$, leading to $R_+^2=0$ and $R_-^2=-2a>0$. The four-velocity reads
\begin{subequations}
\begin{align}
    V_0^\mu&~=~\widetilde{A}^\mu+\tfrac{1}{2}\widetilde{B}^\mu\tau^2+\widetilde{C}^\mu\tau+\widetilde{D}^\mu\cos(R_-\tau)+\widetilde{E}^\mu\sin(R_-\tau)\,,\\
    \widetilde{A}^\mu&~=~\left(1-\frac{\chi_0^2}{R_-^4}(\chi_0^2-\chi_1^2),0,-\frac{\chi_0\chi_1}{R_-^4}(\chi_0^2-\chi_1^2-\chi_2^2),0,-\frac{\chi_0\chi_1\chi_2\chi_3}{R_-^4}\right)\,,\\
    \widetilde{B}^\mu&~=~\left(\frac{\chi_0^2}{R_-^2}(\chi_0^2-\chi_1^2+R_-^2),0,\frac{\chi_0\chi_1}{R_-^2}(\chi_0^2-\chi_1^2-\chi_2^2+R_-^2),0,\frac{\chi_0\chi_1\chi_2\chi_3}{R_-^2}\right)\,,\\
    \widetilde{C}^\mu&~=~\left(0,\frac{\chi_0}{R_-^2}(\chi_0^2-\chi_1^2+R_-^2),0,\frac{\chi_0\chi_1\chi_2}{R_-^2},0\right)\,,\\
    \widetilde{D}^\mu&~=~\left(\frac{\chi_0^2}{R_-^4}(\chi_0^2-\chi_1^2),0,\frac{\chi_0\chi_1}{R_-^4}(\chi_0^2-\chi_1^2-\chi_2^2),0,\frac{\chi_0\chi_1\chi_2\chi_3}{R_-^4}\right)\,,\\
    \widetilde{E}^\mu&~=~\left(0,-\frac{\chi_0}{R_-^3}(\chi_0^2-\chi_1^2),0,-\frac{\chi_0\chi_1\chi_2}{R_-^3},0\right)\,.
\end{align}
\end{subequations}
We proceed as in~\eqref{eqn:: defining Lorentz} to find
\begin{multline}\label{eqn:: IVa semicubical}
    V_0^\mu~=~\left(\sqrt{-\widetilde{A}^\mu \widetilde{A}_\mu}-\frac{1}{2}\frac{\widetilde{A}^\mu \widetilde{B}_\mu}{\sqrt{-\widetilde{A}^\mu \widetilde{A}_\mu}}\tau^2,\sqrt{\widetilde{C}^\mu \widetilde{C}_\mu}\tau,-\frac{1}{2}\frac{\widetilde{A}^\mu \widetilde{B}_\mu}{\sqrt{-\widetilde{A}^\mu \widetilde{A}_\mu}}\tau^2,\right.\\\left.-\sqrt{\widetilde{D}^\mu \widetilde{D}_\mu}\sin(R_-\tau),\sqrt{\widetilde{D}^\mu \widetilde{D}_\mu}\cos(R_-\tau)\right)\,,
\end{multline}
whose spatial profile is the semicubical parabola $y^2=\tfrac{2}{9}\tfrac{(\widetilde{A}^\mu \widetilde{B}_\mu)^2}{(-\widetilde{A}^\nu \widetilde{A}_\nu)(\widetilde{C}^\rho \widetilde{C}_\rho)^3}x^3$ in the $x$\nobreakdash--$y$ plane, combined with circular motion in the $x^3$\nobreakdash--$x^4$ plane. We use $(x,y)$ in place of $(x^1,x^2)$ for clarity.
\newline

\paragraph*{\textbf{Case IVc}~---} $b=\chi_2^2\chi_0^2+\chi_3^2(\chi_0^2-\chi_1^2)<0$. In this case, we have both $a<0$ and $b<0$. Hence, $R_+^2=\sqrt{a^2+b}+a<0$, yet $R_-^2=\sqrt{a^2+b}-a>0$. We then write $a=-\alpha$, $b=-\beta$ such that $R_+^2=-(\alpha-\sqrt{\alpha^2-\beta})=-\rho_-^2$. The four-velocity reads
\begin{subequations}\label{eqn:: Case 5c}
\begin{align}
    V_0^\mu~&=~A^\mu+B^\mu\cos(\rho_-\tau)+C^\mu\sin(\rho_-\tau)+D^\mu\cos(R_-\tau)+E^\mu\sin(R_-\tau)\,,\\
    A^\mu&~=~\left(1-\frac{\chi_0^2}{b}(\chi_2^3+\chi_3^2),0,-\frac{\chi_0\chi_1\chi_3^2}{b},0,-\frac{\chi_0\chi_1\chi_2\chi_3}{b}\right)\,,\\
   B^\mu&~=~\frac{1}{R^2}\left(-\frac{\chi_0^2}{\rho_-^2}(\chi_0^2-\chi_1^2+R_-^2),0,-\frac{\chi_0\chi_1}{\rho_-^2}(\chi_0^2-\chi_1^2-\chi_2^2+R_-^2),0,-\frac{\chi_0\chi_1\chi_2\chi_3}{\rho_-^2}\right)\,, \\
    C^\mu&~=~\frac{1}{R^2}\left(0,\frac{\chi_0}{\rho_-}(\chi_0^2-\chi_1^2+R_-^2),0,\frac{\chi_0\chi_1\chi_2}{\rho_-},0\right)\,,\\
    D^\mu&~=~\frac{1}{R^2}\left(\frac{\chi_0^2}{R_-^2}(\chi_0^2-\chi_1^2+\rho_-^2),0,\frac{\chi_0\chi_1}{R_-^2}(\chi_0^2-\chi_1^2-\chi_2^2+\rho_-^2),0,\frac{\chi_0\chi_1\chi_2\chi_3}{R_-^2}\right)\,,\\
    E^\mu&~=~\frac{1}{R^2}\left(0,-\frac{\chi_0}{R_-}(\chi_0^2-\chi_1^2+\rho_-^2),0,-\frac{\chi_0\chi_1\chi_2}{R_-},0\right)\,,
\end{align}
\end{subequations}where $R^2=R_-^2-\rho_-^2$. Proceeding once more as in~\eqref{eqn:: defining Lorentz}, one may rewrite this four-velocity as
\begin{multline}\label{eqn:: IVa circular}
    V_0^\mu~=~\left(\sqrt{-A^\mu A_\mu},-\sqrt{B^\mu B_\mu}\sin(\rho_-\tau),\sqrt{B^\mu B_\mu}\cos(\rho_-\tau),\right.\\\left.-\sqrt{D^\mu D_\mu}\sin(R_-\tau),\sqrt{D^\mu D_\mu}\cos(R_-\tau)\right)\,,
\end{multline}identifying the trajectory as independent circular motions in the $x^1$\nobreakdash--$x^2$ and $x^3$\nobreakdash--$x^4$ planes.

\bibliography{zzz_bibliography}

\end{document}